\documentclass{statsoc}

\newcommand{\citeg}[1]{(e.g., \citealp{#1})}
\usepackage{amsmath,amssymb,amsbsy,bbm}
\usepackage{graphicx}
\usepackage{booktabs}
\usepackage[tight]{subfigure}
\usepackage{url}
\usepackage{enumerate}
\usepackage{microtype}

\newcommand{\bx}{\mathbf{x}}
\newcommand{\D}{\mathcal{D}}
\newcommand{\N}{\mathcal{N}}

\newcommand{\ave}[2]{{\mathbb E}_{#1}\kern-2pt\left[ #2 \right]}

\newcommand{\tT}{\widetilde{T}}
\newcommand{\dummy}{\hat}

\newcommand{\te}{\!=\!}
\newcommand{\tp}{\!+\!}
\newcommand{\tm}{\!-\!}
\newcommand{\tlt}{\!<\!}
\newcommand{\tgt}{\!>\!}

\newcommand{\la}{\!\leftarrow\!}
\newcommand{\ra}{\!\rightarrow\!}
\newcommand{\tmid}{\!\mid\!}
\newcommand{\g}{\tmid}

\newcommand{\tcdot}{\!\cdot\!}

\newcommand{\pdd}[2]{\frac{\partial #1}{\partial #2}}
\newcommand{\intd}{\mathrm{d}}

\newcommand{\sd}{\pi}

\newcommand{\modone}[1]{#1\,~\mathrm{mod}~1}
\newcommand{\modoneb}[1]{\left(#1\right)\,\mathrm{mod}~1}

\newcommand{\Uniform}{\mathrm{Uniform}}

\title{\hbox{Driving Markov chain Monte Carlo with a} \hbox{dependent random stream}}

\author{Iain Murray}
\address{School of Informatics, University of Edinburgh, UK}
\email{murray@cs.toronto.edu}
\author[I. Murray and L.T. Elliott]{Lloyd T. Elliott}
\address{Gatsby Computational Neuroscience Unit, University College London, UK}
\email{elliott@gatsby.ucl.ac.uk}

\begin{document}

\maketitle

\begin{abstract}
Markov chain Monte Carlo is a widely-used technique for
generating a dependent sequence of samples from complex distributions.
Conventionally, these methods require a source of independent random
variates. Most implementations use pseudo-random numbers instead
because generating true independent variates with a physical system is
not straightforward. In this paper we show how to modify some commonly used
Markov chains to use a dependent stream of random numbers in place of
independent uniform variates. The resulting Markov chains have the
correct invariant distribution without requiring detailed knowledge
of the stream's dependencies or even its marginal distribution. As a
side-effect, sometimes far fewer random numbers are required to obtain
accurate results.
\end{abstract}

\section{Introduction}

Markov chain Monte Carlo (MCMC) is a long-established, widely-used
method for drawing samples from complex distributions
\citeg{neal1993}. Simulating a Markov chain yields a sequence of
states that can be used as dependent samples from a target
distribution of interest. If the initial state of the chain is drawn
from the target distribution, the marginal distribution of every state
in the chain will be equal to the target distribution. It is
usually not possible to draw a sample directly from the target distribution. In that case the
chain can be initialized arbitrarily, and the marginal distributions
of the states will converge to the target distribution as the
simulation progresses. The states of the chain are often used to
approximate expectations, such as those required by stochastic
approximation optimization algorithms, or Bayesian inference.
Estimating expectations does not require independent samples, and any
bias from an arbitrary initialization disappears in the limit of long
Markov chain simulations. Bias can also be reduced by discarding all samples
collected during an initial `burn-in' period, and estimating an expectation using only the samples
collected after a certain point in the chain.

Usually Markov chains are simulated with deterministic computer code
that makes choices based on an independent stream of random variates,
often drawn from $\Uniform[0,1]$. In practice, most implementations
use pseudo-random numbers as a proxy for `real' random numbers. Noisy
physical systems can potentially be used to provide `real' random
numbers, although these are slow when used rigorously (see
section~\ref{sec:rng}). Early hardware implementations of stochastic
neural networks had problems with dependencies in their noise sources
\citep{alspector1989}. Back then, hope and empirical performance
(rather than theory) were used to justify using dependent random
numbers.

Taking a standard MCMC code and replacing its independent
variates with dependent variates will usually lead to biases. As a
simple tractable example, consider the Gaussian random walk
\begin{equation}
    x_{t+1} \;\leftarrow\; \alpha x_t + \sqrt{1-\alpha^2}\; \nu_t,\qquad \nu_t\sim\N(0,1),
\end{equation}
where $0\tlt\alpha\tlt1$. This chain will explore a unit Gaussian
distribution: $x_t$ will be distributed as $\N(0,1)$, either in the
limit $t\ra\infty$, or for all $t$ if $x_1$ was drawn from $\N(0,1)$.
If the noise terms were dependent in some way, then the variance of
the equilibrium distribution could be altered. For example, if the
random variates at even time-steps were a copy of the variate at the
previous time step (i.e., $\nu_2 \te \nu_1$, $\nu_4 \te \nu_3$,
\dots)\ then the equilibrium variance would increase from~1 to
$(1+\alpha)^2/(1+\alpha^2)$. In most MCMC algorithms the random
variates driving the Markov chain\,---\,$\nu$ in this
example\,---\,must be independent and come from a specified marginal
distribution.

Given that the output of an MCMC algorithm is a dependent sequence, it
is not clear that the variates used to drive the chain must
necessarily be independent.
In this paper we give a constructive demonstration that it is possible
to modify many standard MCMC codes so that they can be driven by a
dependent random sequence. Surprisingly, our construction does not
depend on the functional form of the dependencies, or on the marginal
distribution of the dependent stream.

\section{Random number generation in practice}
\label{sec:rng}

Recent papers using MCMC methods rarely discuss or specify the
underlying random number generator used to perform Markov chain
simulations. A common view is that many pseudo-random number
generators are good enough for many purposes, but only some of the
slower more complex generators are completely trusted. For example,
Skilling (personal communication) reports his experience on using one
of the generators from Numerical Recipes \citep{press2002} within
BayeSys \citep{skilling2004}:
\begin{quote}
    \textsl{\llap{``}Although I used this procedure
    heavily for more than ten years without noticing any damage, one recent
    test
    gave weakly but reproducibly biased results. This bias vanished on
    replacing the generator, so I have discarded \texttt{ran3} despite its
    speed.''}
\end{quote}
Similarly there are known problems with the default, although fast,
random number generator provided by Matlab \citep{savicky2006}.
Switching to more modern generators is easy to do, and it is widely
thought that the state-of-the-art pseudo-random number generators are
good enough for most Monte Carlo purposes. Nevertheless, it is hard to
be certain there will not be problems with using pseudo-random numbers
in a given application.

The ``Flexible Bayesian Modelling'' (FBM) software by
\citet{neal1995a} calls a standard Unix pseudo-random number generator
known as \texttt{drand48}. Although this pseudo-random number
generator performs reasonably, it has known problems in some
circumstances \citep{entacher1998}, which (in a seemingly isolated
incident) have lead to errors in an application~\citep{gartner2000}.
Although FBM would probably have no practical problems using
\texttt{drand48} only, the code is somewhat robust to possible difficulties
by also making use of a file of 100,000 `real' random numbers.

Combining numbers from a physical source and a pseudo-random generator
is also common in cryptography. For example the Intel generator works
in this way~\citep{jun1999}. However, most of the development of these
methods has been focussed on ensuring that the random bits are
unpredictable enough to prevent encryption schemes from being cracked,
and also that the discovery of the internal state of the generator
does not allow the bits to be easily reconstructed.
The Intel team were careful to be quite guarded about the quality of
the random numbers from their hardware. Some deviations from
independent uniform randomness are possible, but no analysis of the
resulting potential bias in Monte Carlo work has been reported.

From a computer science theory perspective, obtaining independent random bits from
a random source is known as `randomness extraction'
\citeg{shaltiel2002}, and can come with a variety of guarantees.
In practice, generating independent uniform random numbers from a
physical source is tricky and slow. See the HotBits online service
(\url{http://www.fourmilab.ch/hotbits/})
for an example of what can be involved.
The Linux random number generators
(\texttt{/dev/random} and \texttt{/dev/urandom}) operate by gleaning
entropy from system events. But their entropy pool can be exhausted quickly \citep{gutterman2006}.

Random number generating hardware could potentially be made simpler
and faster if the requirement for independent uniform variates were
removed. We will show that it is indeed possible to adapt Markov chain
Monte Carlo to use any source of random numbers, and that far less
randomness may be required than is commonly used.

\section{Markov chain Monte Carlo, setup and preliminaries}
\label{sec:mcmcbackground}

A Markov chain transition operator $T(x'\la x)$ specifies a
probability distribution over the next state~$x'$ given the current
state~$x$.
If the Markov chain is defined over a finite set of states, then
$T(x'\la x)$~is a probability mass function over $x'$ for every fixed~$x$.
Otherwise, if the states are continuous, then $T(x'\la x)$ is a
probability density. Given a target distribution~$\sd(x)$, standard
MCMC techniques require a transition operator that leaves the target
\emph{invariant}:
\begin{equation}
    \sd(x') = \sum_x T(x'\la x)\, \sd(x)
    \qquad\text{or}\qquad \sd(x') = \int T(x'\la x)\, \sd(x) \;\intd{x}.
    \label{eqn:invariant}
\end{equation}
The Markov chain produced by the operator should also be
\emph{ergodic}, meaning that any starting state should eventually be
`forgotten' and the marginal distribution of the chain should converge
to the target distribution,~$\sd(x)$. Such ergodic chains have the
target distribution as their \emph{equilibrium distribution}. A simple
sufficient condition for ergodicity in the discrete setting is that
there is a non-zero probability of transitioning from any state to any
other state in $K$ transitions for some finite~$K$ \citeg{neal1993}.
Some care is required for defining ergodicity in the continuous
setting.
The technical details, and more formal definitions of the above, are outlined by
\citet{tierney1994a}.

A transition operator that leaves the target distribution invariant
(one that satisfies equation~\ref{eqn:invariant}) can be created by
concatenating together several operators that each
satisfy~\eqref{eqn:invariant} individually. The component operators do
not need to be ergodic in isolation. To build an MCMC method, first
one finds a useful set of operators that each
satisfy~\eqref{eqn:invariant}, and second, one checks ergodicity for
the composition of the operators \citeg{tierney1994a}.

For any transition operator satisfying~\eqref{eqn:invariant}, one can
define a
\emph{reverse operator}:
\begin{equation}
    \tT(x\la x') =
    \frac{T(x'\la x)\,\sd(x)}{\int T(x'\la x)\,\sd(x)\;\intd{x}} =
        \frac{T(x'\la x)\,\sd(x)}{\sd(x')} \,.
    \label{eqn:reverseoperator}
\end{equation}
The existence of such a reverse operator is a necessary condition for
any operator that leaves the target distribution invariant. This
condition is also sufficient: if the reverse operator exists, then we
can form a symmetric version of (\ref{eqn:reverseoperator}),
\begin{equation}
    \tT(x\la x')\,\sd(x') = T(x'\la x)\,\sd(x),
    \label{eqn:genbalance}
\end{equation}
which, integrated or summed over $x$ on both sides, implies~\eqref{eqn:invariant}.
If
an operator is \emph{reversible}, $T\te\tT$, the condition is known as
\emph{detailed balance}.

In many MCMC algorithms involving distributions on random vectors, the
coordinates of the vector are updated sequentially by component
operators of the algorithm. For such algorithms, we need only consider
operators that act on one-dimensional random variables. Suppose $T$ is
a component operator of an MCMC algorithm. Often $T$ is implemented as
a deterministic function of a uniform variate $u$, $x'\te\tau(u; x)$.
In this case, the Markov chain can be driven using a sequence of
uniform random variates $u_1,u_2,\ldots$, by initializing the chain at
$x_1$ and computing $x_2 \te \tau(u_1; x_1),x_3 \te \tau(u_2,x_2),\ldots$.

For a transition operator that acts on a continuous random
variable, we can take $\tau$ to be the inverse cumulative distribution
function (cdf) of the transition operator:
\begin{equation}
    \tau^{-1}(x';x) = \int_{-\infty}^{x'} T(\dummy{x}\la x) \;\intd\dummy{x}.
    \label{eqn:cumulative}
\end{equation}
Evaluating the inverse cdf, $\tau(\,\cdot\,;x)$, at a uniform
variate~$u$ produces a sample from the transition operator. If an
independent uniform variate~$u$ is drawn for each update, then the
Markov chain will have the desired equilibrium distribution.

For discrete variables
\begin{equation}
    \tau(u;x) = \min \left\{ x' : u > \sum_{\dummy{x}<x'} T(\dummy{x}\la x) \right\},
\end{equation}
is the analogous deterministic function.  We can implement
the reverse transition operator as a deterministic function of a
uniform variate by setting $x\te\tilde{\tau}(u;x')$ where
\begin{equation}
    \tilde{\tau}^{-1}(x;x') = \int_{-\infty}^{x} \tT(\dummy{x}\la x') \;
\intd\dummy{x}.
\end{equation}
The reverse move for discrete variables is defined analogously.

\section{Constructing a Markov chain from a dependent stream}
\label{sec:construction}

We now assume that we do not have access to independent random
variates with which to drive the chain. Instead, we have access to a
stream of random variates, $d_t$, with arbitrary dependencies.
We will find that it is possible to construct valid Markov chain Monte
Carlo algorithms\,---\,chains with the correct invariant
distribution\,---\,without knowing any details of the dependency
structure of the stream, or even the marginal distribution of its
output. The speed with which the chain reaches equilibrium will depend
on properties of the stream. Proving ergodicity will require some
additional, but weak, conditions.

We wish to adapt existing MCMC algorithms to use dependent streams. We
initially restrict attention to operators driven by a single uniform
random variate, as outlined in section~\ref{sec:mcmcbackground}. Given
that we do not have an external source of uniform variates, we will
include one in our state; we will run a Markov chain on a pair~$(x,u)$
with target distribution
\begin{equation}
    \sd_a(x,u) =
    \begin{cases}
        \sd(x) & \text{if~~$0<u<1$} \\
        0 & \text{otherwise.} \\
    \end{cases}
    \label{eqn:aux_target}
\end{equation}
To obtain samples from the target distribution we can simulate a
Markov chain that has equilibrium distribution~\eqref{eqn:aux_target},
and discard the auxiliary~$u$ samples. Our Markov chain Monte Carlo
algorithm will concatenate transition operators that each leave the
joint auxiliary distribution invariant.

\subsection{Operator 1}

We define an operator that adds the current stream output, $d_t$, to
the uniform variate with a wraparound at one:
\begin{equation}
    u_{t+1} = \modone{(u_t + d_t)}.
\end{equation}
By $u\;\mathrm{mod}\;1$ we mean $u \tm \lfloor u\rfloor$
(e.g., $\modone{9.1} = 0.1$~~and~~$\modone{-8.3} = 0.7$). This
operation leaves the target uniform distribution invariant: the
marginal distribution of the sum modulo one of any real number and a
uniform draw is $\Uniform[0,1]$.

\subsection{Operator~2}

Given a sample $(x,u)$ from the auxiliary distribution~\ref{eqn:aux_target},
we could obtain another dependent sample from the target distribution,
$\sd(x)$, by using the uniform variate to take a Markov chain step:
\begin{equation}
    x' = \tau(u;x).
    \label{eqn:normal_update}
\end{equation}
The function $\tau$ is the inverse of the cumulative distribution
function~\eqref{eqn:cumulative} for the transition operator of the
MCMC algorithm that we are adapting. To create an update that leaves
the joint auxiliary distribution invariant we must also modify the
uniform variate such that a reverse procedure exists that will map
$(x',u')\ra(x,u)$. Such a reverse procedure must exist if we wish to
satisfy the balance condition~\eqref{eqn:genbalance} for the joint
auxiliary distribution.

For continuous variables we simply set the uniform variate to the
value that would drive the reverse transition operator to move
from~${x'\ra x}$:
\begin{equation}
    u' = \int_{-\infty}^{x} \tT(\dummy{x}\la x')\;\intd\dummy{x}.
    \label{eqn:operator1}
\end{equation}
The new $(x',u')$ pair is the result of the two changes of variable in
equations~\eqref{eqn:normal_update} and~\eqref{eqn:operator1}.
Therefore, the joint
distribution of $(x', u')$ can be found by multiplying $\sd_a(x, u)$
by the Jacobian of the change of variables:
\begin{align}
    P(x', u') &= \sd_a(x,u)\cdot
    \left|\pdd{u}{x'}\right|\cdot
    \left|\pdd{x}{u'}\right|\\
    &= \sd(x)\cdot
    T(x'\la x)\cdot
    1/ \tT(x\la x')\\
    &= \sd(x').
\end{align}
We substituted \eqref{eqn:reverseoperator} and \eqref{eqn:aux_target},
assuming that the initial state is feasible (i.e., that $u\in[0,1]$).
By construction, $u'\in[0,1]$ and so the probability of the final pair
is given by the target distribution, $\sd_a(x',u')$. Thus $\sd_a$, the
target distribution \eqref{eqn:aux_target}, is invariant under this
operator.

For discrete variables there are a range of uniform values,
$u\in[u_{\mathrm{min}},u_{\mathrm{max}}]$, that yield the same forwards
move $x' \te \tau(u;x)$. Similarly a reverse move, $x \te
\tilde{\tau}(u';x')$, defined with the reverse transition operator
could be made using any value in the range
$u'\in[u'_{\mathrm{min}},u'_{\mathrm{max}}]$. The ranges are given by:
\begin{align}
    u_{\mathrm{min}} &= \sum_{\dummy{x}<x'} T(\dummy{x}\la x) &
    u'_{\mathrm{min}} &= \sum_{\dummy{x}<x} \tT(\dummy{x}\la x') \notag \\
    u_{\mathrm{max}} &= u_{\mathrm{min}} + T(x'\la x) &
    u'_{\mathrm{max}} &= u'_{\mathrm{min}} + \tT(x\la x').
\end{align}
We need a procedure that picks a new uniform variate~$u'$ in a way
that could be reversed to recover the original variate~$u$.
We choose the procedure that leaves the fraction through the available ranges constant:
\begin{align}
    u' &= u'_{\mathrm{min}} +
    (u'_{\mathrm{max}}-u'_{\mathrm{min}})
    \frac{u-u_{\mathrm{min}}}{u_{\mathrm{max}}-u_{\mathrm{min}}} \\
    &=
    \sum_{\dummy{x}<x} \tT(\dummy{x}\la x') +
    \frac{\tT(x\la x')}{T(x'\la x)}
    \bigg(u - \sum_{\dummy{x}<x'} T(\dummy{x}\la x)\bigg)
    .
    \label{eqn:finduprime}
\end{align}
In the discrete setting, the probability of the state
update~\eqref{eqn:normal_update} is one. The probability of obtaining
a new joint state, deterministically transformed from an equilibrium
state, is:
\begin{equation}
    P(x',u') = \sd(x,u) \cdot 1 \cdot \left|\pdd{u}{u'}\right|
        = \sd(x) \frac{T(x'\la x)}{\tT(x\la x')}
        = \sd(x').
\end{equation}
Again the update leaves the joint auxiliary distribution invariant.

\subsection{The algorithm and its properties}

Applying Operators~1 and~2 alternately in sequence will leave the
joint auxiliary distribution invariant. If the transition operator of
the MCMC algorithm that we are adapting is formed by a composition of
transition operators, then the underlying transition distribution~$T$
used in Operator~2 should cycle through each component of the
composition. The algorithm is presented in Fig.~\ref{alg:new_method}.

Although the sequence resulting from this algorithm will always leave
the auxiliary distribution invariant, it will only be ergodic if the
random stream satisfies some conditions.  For example, if the stream
were to always emit a zero, then Operator~1 will have no effect. In
this case, if $T$ were given by a single reversible component, then
the sequence would alternate between $x_1$ and $\tau(0, x_1)$ and
would thus not be ergodic.
At the other extreme, if the stream is actually an independent source
of $\Uniform[0,1]$ random variates, then $u'\te\modoneb{u + d}$ is also a
uniform variate that is independent of the current Markov chain state
$(u,x)$. In this case, the statistics of the original Markov chain are
recovered exactly.

If the original transition operator that we are modifying is ergodic, it will have sets~$X'$ that can be
reached with positive probability from any given state~$x$ in
a finite number of steps. The transitions to a set $X'$ can
be realized by a set of sequences of driving random numbers. If every
sequence of stream outputs can be realized
(i.e., if the density of~$d_t$ dominates the Uniform$[0,1]$
distribution for every possible stream history), then our method will
have the same reachable sets. Thus, the proofs of ergodicity for most
algorithms will be maintained. Our method may also produce ergodic
chains for some more constrained dependent streams, although a proof
would depend on the MCMC algorithm being modified.

If ideal independent random numbers are available, but are slow to
obtain, these could be used for a small fraction of iterations and a
pseudo-random stream used for the remaining iterations. The resulting
chain will leave the target distribution invariant, and under weak
conditions it will also be ergodic.
For example, the usual proofs of ergodicity rely on sets of states
that have positive probability measure after $K$ transition steps,
regardless of starting position, for some finite~$K$. These sets are
still accessible if we regularly perform~$K$ of the original
transitions, using independent Uniform$[0,1]$ numbers. In between
these sequences of updates we can interleave long runs of transitions
driven by the dependent stream. These intermediate runs might each be
constrained to some partition of the space, but the combined chain can
potentially get anywhere and will still converge.
In contrast, ad-hoc schemes for combining poor-quality random
numbers with an ideal physical source provide no such guarantees.

\begin{figure}
\begin{tabular}{l}
\toprule
    \hbox{Inputs: }Initial augmented state $(x,u)$; dependent stream $\D$; operators $\{T_n(x'\la x)\}_{n=1}^{N}$\rlap{,}\\
    \phantom{\hbox{Inputs: }}with
    inverse cumulatives $\tau_n(u;x)$
    and reverse operators $\{\tT_n(x'\la x)\}$
    (see~\eqref{eqn:reverseoperator})
    \\
    \vspace*{-.4cm}\\
    \midrule
\begin{minipage}{0.9\linewidth}
\begin{enumerate}[1.]
\setlength{\itemsep}{2pt}
\setlength{\parskip}{0pt}
\setlength{\parsep}{0pt}
\setlength{\leftmargin}{0pt}
\item \textbf{for} $n = 1\dots N$\textbf{:}
    \vspace*{0.02in}
\item[] \qquad\textit{\small Operator 1:}
\item \qquad Query the stream: $d\sim\D$
\item \qquad $u \leftarrow \modoneb{u + d}$ \qquad \textit{\small --- introduces randomness, replaces $u\sim\Uniform[0,1]$}
    \vspace*{0.05in}
\item[] \qquad\textit{\small Operator 2:}
\item \qquad $x_\mathrm{old} \leftarrow x$
\item \qquad $x \leftarrow \tau_n(u;x)$ \qquad\qquad\quad \textit{\small --- as in conventional update with random uniform $u$}
\item \qquad \textbf{if} $x$ is discrete\textbf{:}
\item \qquad\qquad $u \leftarrow
    \sum_{\dummy{x}<x_\mathrm{old}} \tT(\dummy{x}\la x) +
    \frac{\tT(x_\mathrm{old}\,\la\, x)}{T(x\,\la\, x_\mathrm{old})}
    \Big(u - \sum_{\dummy{x}<x} T(\dummy{x}\la x_\mathrm{old})\Big)$
    \hspace*\fill\eqref{eqn:finduprime}
\item \qquad \textbf{else:}
\item \qquad \qquad $u \leftarrow \int_{-\infty}^{x_\mathrm{old}} \tT_n(\dummy{x}\la x)\;\intd\dummy{x}.$
    \hspace*\fill\eqref{eqn:cumulative}
\end{enumerate}
\end{minipage}\\
\bottomrule
\end{tabular}
\caption{Procedure for applying a set of simple conventional Markov chain
transition operators while using a dependent stream.
Section~\ref{sec:converting} discusses applicability to less-tractable
transition operators.}
\label{alg:new_method}
\end{figure}

\section{Application to common MCMC algorithms}
\label{sec:converting}

Not all commonly-used MCMC procedures have transition operators that
use a single driving random number~$u$, and that are tractable enough
to apply the method described in the previous section. However, if a
method can be implemented in practice, it can usually be broken down
into a series of deterministic operations driven by a set of uniform
variates. Also, a corresponding reverse operator will have been
identified to prove that the necessary balance
condition~\eqref{eqn:genbalance} holds. These observations allow us
to generalize our procedure (Fig.~\ref{alg:new_method}) to other
algorithms.

Our strategy for adapting existing Markov chain Monte Carlo algorithms
to use a dependent stream will be as follows: 1)~include the required
random numbers in the state; 2)~update uniform variates by addition
module-one with the stream; 3)~apply deterministic moves that change
the variables of interest using the uniform variates, and update the
uniform variates so that the move would be reversed under the reverse
operator. We now give some details of applying this strategy to some
of the most commonly-used MCMC algorithms.

\subsection{Gibbs Sampling}

Gibbs sampling is one of the most commonly used MCMC methods for
sampling a set of variables~$x \te [x_1,\dots, x_N]$. Each variable is
resampled in turn from its conditional distribution. For continuous
variables:
\begin{equation}
    T_n(x'\la x) = \sd(x'_n\g x_{m\neq n})\, \delta(x'_{m\neq n} - x_{m\neq n}),
\end{equation}
and for discrete:
\begin{equation}
    T_n(x'\la x) =
    \begin{cases}
        \sd(x'_n\g x_{m\neq n}) & x'_{m\neq n} = x_{m\neq n} \\
        0 & \text{otherwise.}
    \end{cases}
\end{equation}
Provided these conditional distributions are sufficiently tractable,
the algorithm in Fig.~\ref{alg:new_method} applies directly.
Example code is provided as supplementary material.

Gibbs sampling is also performed on distributions without tractable
conditional distributions. The techniques for sampling from these
distributions often require a random number of independent uniform
variates \citeg{gilks1992,gilks1992a,devroye1986}. An example
application to a method with this flavor is given in
Section~\ref{sec:slice}.

\subsection{Metropolis--Hastings}
In the Metropolis--Hastings transition rule \citep{hastings1970}, a
sample is drawn from an arbitrary proposal density $q(x';x)$. The proposal is
accepted with probability
\begin{equation}
    a(x';x) = \min\left(1,\;\frac{q(x;x')\,\sd(x')}{q(x';x)\,\sd(x)}\right),
\end{equation}
otherwise the proposal is rejected and the Markov chain state stays
fixed for a time step. The method is often implemented by drawing a
$\Uniform[0,1]$ variate, $u_a$, and accepting if it is less than the
acceptance probability.

We consider cases where the proposal is implemented by drawing a
uniform variate, $u_q$, and using the inverse cdf to compute the
proposal $x'$ that satisfies
\begin{equation}
    u_q = \int_{-\infty}^{x'} q(\tilde{x}; x)\;\intd\tilde{x}.
    \label{eqn:propose}
\end{equation}
As in section~\ref{sec:construction}, we include the uniform variates
in our Markov chain state. In this augmented state space the target
equilibrium probability of the triple $(x,u_q,u_a)$ is
$\sd(x)\tcdot1\tcdot1$ or zero if the uniform variates are
outside~$[0,1]$. As before, the uniform variates can be updated by
addition of the stream output modulo-one.

The triple can be deterministically transformed with three changes of
variable: $u_q\ra x'$, $u_a\ra u_a'$, $x\ra u_q'$. First $u_q$ is used
to make the proposal $x'$ satisfying Equation~\eqref{eqn:propose}. If
the proposal is rejected (i.e., $u_a\tgt a(x';x)$), then the whole
Markov chain state $(x,u_a,u_q)$ will be left unchanged. If the
proposal is accepted we continue by creating
\begin{equation}
    u_a' = \frac{u_a}{a(x';x)}\,a(x;x') = u_a\,\frac{\sd(x)\,q(x';x)}{\sd(x')q(x;x')},
\end{equation}
a variate that would accept the reverse move $x'\ra x$, with $u_a'$ the
same fraction through the acceptable range as $u_a$ was for the
forward transition. Finally, we create
\begin{equation}
    u_q' = \int_{-\infty}^{x} q(\tilde{x}; x')\;\intd\tilde{x},
    \label{eqn:proposeback}
\end{equation}
the uniform variate that would propose the original state from the new state.

The probability of a new state is:
\begin{equation}
\begin{split}
    p(x',u_q',u_a') &=
    \pi(x,u_q,u_a)\cdot
    \left|\pdd{u_q}{x'}\right|\cdot
    \left|\pdd{u_a}{x_{a'}}\right|\cdot
    \left|\pdd{x}{u_{q'}}\right|\\
    &= \sd(x)\cdot q(x';x)\cdot \frac{\sd(x')\,q(x;x')}{\sd(x)\,q(x';x)}\cdot \frac{1}{q(x;x')}\\
    &= \sd(x') = \sd(x',u_q',u_a'),
\end{split}
\end{equation}
the target invariant distribution. Our version of Metropolis--Hastings
with a simple proposal distribution could be used to sample target
distributions with arbitrary probability density
functions. Therefore, sampling using a dependent random stream is always
possible in principle.

\subsection{More complex operators: slice sampling}
\label{sec:slice}

Slice sampling \citep{neal2003a} is a family of methods that combine
an auxiliary variable sampler with an adaptive search procedure. The
technical requirements are similar to Metropolis--Hastings
methods, yet slice sampling tends to be easier to use as it is more
robust to any initial choices of step-size parameters. We use slice
sampling as a case study of a method that requires a random number of
uniform variates within an update.

A slice sampler augmented for use with dependent streams is given in Fig.~\ref{alg:slice}. Only a
few additions have been made to the standard ``linear stepping out'' slice sampling algorithm
by \citet{neal2003a}. As with standard
slice sampling, the code only needs to evaluate an unnormalized version
of the target distribution at a set of points.

An arbitrary number of random draws are used during a standard slice sampling update.
We must augment the state space with all of the variates used in
an update in order to ensure that the reverse operator satisfies the balance
condition~\eqref{eqn:genbalance}. We would like our augmented state space to
be of finite dimension, and so we choose a finite~$K$ and augment our state space with
$K$ variates with independent Uniform$[0,1]$ target distributions.
Unlike in standard slice sampling,
our augmented version might `reject' a sample. This will happen if
if all $K$
variates are exhausted before an acceptable move is found. In that case, our
slice sampling iteration gives up and the chain stays fixed for a time step. Including these rejections leaves
the distribution invariant.

\begin{figure}
\begin{tabular}{l}
\toprule
    \rlap{Inputs: }\phantom{\hbox{Output: }}Initial augmented state $(x,u_1,u_2,\dots,u_K)$; dependent stream $\D$;\\
    \phantom{\hbox{Output: }}unnormalized target distribution $\sd^*(x) = Z\sd(x)$; step-size $w$
    \\
    \rlap{Output:}\phantom{\hbox{Output: }}Updated augmented state.\\
    \vspace*{-.4cm}\\
    \midrule
\begin{minipage}{0.8\linewidth}
\begin{enumerate}[1.]
\setlength{\itemsep}{2pt}
\setlength{\parskip}{0pt}
\setlength{\parsep}{0pt}
\setlength{\leftmargin}{0pt}
\item[] \textit{\small Set target probability threshold:}
\item $u_1 \leftarrow \modoneb{u_1 +d},\quad d\sim\D$ \label{alg:u1}
\item $y = u_1 \sd^*(x)$
\item[] \textit{\small Initialize bracket on which to make proposals:}
\item $u_2 \leftarrow \modoneb{u_2 + d},\quad d\sim\D$ \label{alg:u2}
\item $x_{L,1} \leftarrow x - u_2w$
\item $[x_L,x_R] \leftarrow [x_{L,1},\,x_{L,1}\tp w]$
\item \textbf{while} $(\sd^*(x_L) > y)$\textbf{:}
\item \qquad $x_L = x_L - w$
\item \textbf{while} $(\sd^*(x_R) > y)$\textbf{:}
\item \qquad $x_R = x_R + w$
\item[] \textit{\small Bookkeeping for which uniform variate to use (exit if none are left):}
\item $k \leftarrow 2$
\item $k \leftarrow k + 1$ \label{alg:slice_propose}
\item \textbf{if} $k > K$\textbf{:} \label{alg:bailout1}
\item \qquad \textbf{return} $(x,u_1,u_2,\dots u_K)$\label{alg:bailout2}
\item[] \textit{\small Make proposal on $[x_L,x_R]$ bracket}
\item $u_k = \modoneb{u_k + d},\quad d\sim\D$ \label{alg:uk}
\item $x' \leftarrow x_L + u_k(x_R - x_L)$
\item \textbf{if} $\sd^*(x') < y$\textbf{:}
\item[] \qquad \textit{\small Proposal rejected. Shrink bracket and try again:}
\item \qquad \textbf{if} $x' > x$\textbf{:} $x_R \leftarrow x'$~~\textbf{else:} $x_L \leftarrow x'$
\item \qquad \textbf{goto}~\ref{alg:slice_propose}
\item[] \textit{\small Proposal accepted. Set uniform variates that could reverse this iteration:}
\item $u_1 \leftarrow y / \sd^*(x')$ \label{alg:book1}
\item $u_2 \leftarrow \modone{(x' - x_{L,1})/w}$
\item $u_k \leftarrow (x-x_L)/(x_R - x_L)$ \label{alg:book3}
\item \textbf{return} $(x',u_1,u_2,\dots,u_K)$
\end{enumerate}
\end{minipage}\\
\bottomrule
\end{tabular}
\caption{A univariate slice sampler with ``linear stepping out''
\citep{neal2003a} adapted to use a dependent stream. The differences
from the original algorithm are: the $u_k$ were previously drawn
independently from $\Uniform[0,1]$ (steps~\ref{alg:u1},~\ref{alg:u2}
and~\ref{alg:uk}); the $u_k$ values are now returned
and steps~\ref{alg:book1}--\ref{alg:book3} are new; although
valid, there is usually no reason to limit~$k$,
steps~\ref{alg:bailout1}--\ref{alg:bailout2} are new.}
\label{alg:slice}
\end{figure}

The algorithm in Fig.~\ref{alg:slice} provides a deterministic rule
for advancing from $(x,u_1,\ldots,u_K)$ to $(x',u_1',\ldots,u_K')$.
The final updates to the auxiliary $u$ variables
(steps~\ref{alg:book1}--\ref{alg:book3}) were chosen by identifying a
reverse algorithm that would reverse the move. Our chosen reverse
algorithm first returns the variable of interest, $x'\rightarrow x$,
by running the forwards algorithm without the random stream updates
(omitting steps \ref{alg:u1},~\ref{alg:u2}, and \ref{alg:uk}). The
original auxiliary variables are then recovered by running the
dependent stream $\mathcal{D}$ in time-reverse order and updating
$u_k,u_{k-1},\dots, u_1$ to their previous values (with $u\leftarrow
\modone{u\tm d}$).

We verify that the target distribution,
$\sd(x)$, is left invariant by the algorithm in Fig.~\ref{alg:slice},
by computing the Jacobian of the change of variables implied by the
algorithm:
\begin{align}
    P(x',u_1', u_2', \dots,u_K') &= \sd(x)\cdot
    \left| \pdd{u_1}{y}\right|\cdot
    \left| \pdd{u_2}{x_{L,1}}\right|\cdot
    \left| \pdd{u_k}{x'}\right|\cdot
    \left| \pdd{y}{u_1'}\right|\cdot
    \left| \pdd{x_{L,1}}{u_2'}\right|\cdot
    \left| \pdd{x}{u_k'}\right|\\
    &= \sd(x)\cdot \frac{1}{\sd^*(x)}\cdot \frac{1}{w}\cdot
    \frac{1}{x_R-x_L}\cdot
    \sd^*(x')\cdot w\cdot (x_R-x_L)\\
    &= \sd(x').
\end{align}
The variables $x_{L,1},x_L,x_R$ are initial and final endpoints of the
slice sampling bracket (these are defined in Fig.~\ref{alg:slice}),
and $\sd^*(x)$ is an unnormalized version of the target distribution.

We have demonstrated that it is possible to adapt an MCMC method that
uses a variable number of uniform variates per iteration for use
with dependent streams. Another
possible complication for Markov chain algorithms is a variable state
space size, such as in Reversible Jump MCMC \citep{green1995}, or
samplers for non-parametric Bayesian methods \citeg{neal2000b}.
The potentially unbounded state space in these samplers does not
necessarily introduce any additional problems. As long as the number
of uniform variates required at each iteration can be bounded, it is
possible to use a dependent stream using the techniques developed
in this section.

\section{Effects of dependencies}

Although perfectly independent and uniform samples are difficult to
produce, streams that ostensibly have this property are easily
produced by pseudo-random number generators. Our algorithm will have
the same behavior as standard MCMC in these circumstances, although it
will be a constant factor more expensive due to additional
bookkeeping. Using a stream with strong dependencies will alter the
convergence properties of the chain. In this case, our method will
always leave the target distribution invariant, whereas conventional
methods will usually not.

\subsection{Practical demonstration: the funnel distribution}

We tested our approach on an example target distribution explored in
\citep{neal2003a}. This setup is designed to reflect some of the
realistic challenges found in statistical problems. For simplicity, in
what follows we used a pseudo-random number generator to simulate
the dependent streams.
Our code is available as supplementary material, and the
algorithm used for the pseudo-random number generator
can be easily changed.

The target `funnel' distribution is defined over a ten-dimensional
vector $[v,\bx]$. The first component, $v$, has a Gaussian marginal
distribution, $v\!\sim\!\N(0,3^2)$. The conditional log-variance of
the remaining 9 components is specified by $v$:
$x_i\!\sim\!\N(0,e^v),~i\te1\ldots9$. It is trivially possible to
sample from this distribution exactly by first sampling $v$ and then
sampling each of the~$x_i$. Instead, we used MCMC methods to update
each variable in turn, as is often the situation in real applications.
It is hard to `mix' across the support of this distribution using
simple Metropolis methods, but slice sampling alleviates some of this
difficulty \citep{neal2003a}. In our experiments we used slice
sampling procedures to explore the effect of dependent streams in MCMC\@.
 We initialised our Markov chains with a
draw from the true distribution to simplify the discussion (although
running with an arbitrary initialization and discarding some initial
`burn-in' samples does not significantly change the results).

Consider a `sticky' stream of random variates $\D_p$, where each draw
is a copy of the previous value with `sticking probability'~$p$, and
is otherwise an independent Uniform$[0,1]$ variate. We compared
simulations with two slice samplers, driven by the dependent stream
$\D_p$: 1)~our proposed slice sampler, Fig.~\ref{alg:slice}, with
$K\te10$, 2)~an `incorrect' sampler that uses the usual slice sampling
algorithm, but replaces independent uniform variates with draws from
the dependent stream. We explored a range of dependent streams with
different sticking probabilities,~$p$. When $p\te0$, $\D_p$ produces
independent uniform variates, and the algorithms revert to their
conventional MCMC counterparts.

\begin{figure}
    \centerline{\includegraphics[width=0.8\linewidth]{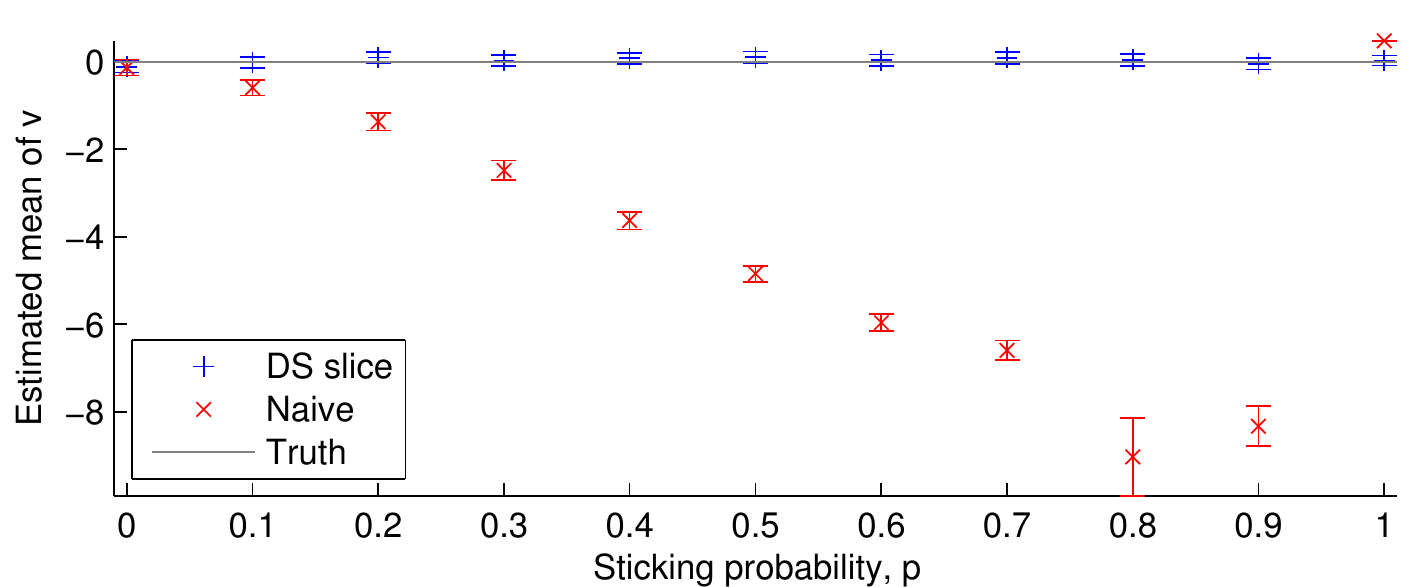}}
    \caption{Estimates of the mean of $v$ from the funnel distribution
    (the true mean is 0) when using slice samplers driven by a
    dependent, `sticky' stream of random variates. The error bars show
    $\pm$2~standard errors, using the effective sample size estimated
    by R-CODA \citep{plummer2006}. The naive approach uses existing
    sampling code and treats the stream as if it gave independent
    uniform outputs, which leads to bias. Our `DS~slice' sampler,
    adapted to dependent streams (Fig.~\ref{alg:slice}), theoretically
    gives no bias but could mix poorly. In practice, the empirical
    diagnostics find that our method works equally well (for this
    task) no matter how sticky the dependent stream is.
    }\label{fig:funnel_biases}
\end{figure}

Following the setup of \citet{neal2003a}, we ran simulations for
240,000 single-variable slice-sampling updates of each variable.
Fig.~\ref{fig:funnel_biases} shows the estimated means of the
log-variance~$v$. When the source stream of random variates never
sticks (i.e., when $p\te0$) the estimates of the mean are consistent with the
true value of zero. The naive use of the dependent stream in a
standard slice sampling code rapidly breaks down, returning
confidently wrong answers for~$p\tgt0$.

\begin{figure}
    \centerline{\includegraphics[width=\linewidth]{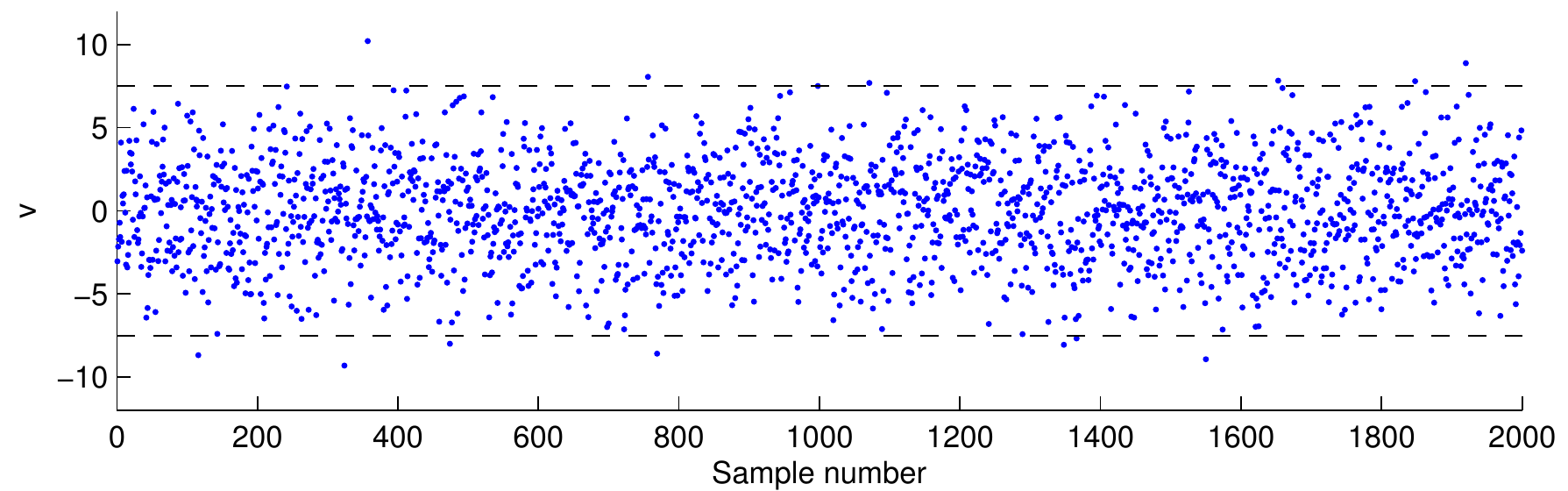}}
    \caption{This figure shows samples of $v$ from the funnel
    distribution, taken after every 120 slice sampling updates of
    every component of the distribution, to match the bottom pane of
    Neal's~(2003) Figure~14. We used our dependent stream slice
    sampler (Fig.~\ref{alg:slice}), using the sticky stream with
    $p\te1$. That is, after the initialization, no external randomness
    is added. The pseudo-randomness here is entirely due to the
    complexity of the slice sampling algorithm itself.
    }\label{fig:funnel_eyeball}
\end{figure}

Our slice sampler, modified for use with the dependent stream by the
procedure described in section \ref{sec:slice}, continues to provide
good estimates of the mean for the sticky streams. In fact, we obtain
good estimates even when the stream always returns a constant and the
chain is deterministic (i.e., when $p\te0$). The behavior of trace
plots, such as Fig.~\ref{fig:funnel_eyeball}, is indistinguishable
from those originally reported in \citet{neal2003a}.

This trace plot (Fig.~\ref{fig:funnel_eyeball}) suggests that, for the
funnel distribution, the interactions between the updates for each
component are complicated enough to explore the distribution without
the need for any external randomness.
In this deterministic limit, only a countable set of points can be
reached by the chain for a given initial condition, and so the chain
is not Harris recurrent.
Its behavior, however, appears to be indistinguishable from the ergodic chains
with~$p\tgt0$.

\subsection{Dependent streams can improve MCMC algorithms}
\label{sec:advantage}

Dependencies in the noise source used to drive an MCMC algorithm
are not necessarily just a problem to
be avoided. Introducing dependencies can actually improve MCMC
performance. As an example, we consider the artificial
problem of sampling
uniformly from the integers modulo~100. By symmetry, applying the
following update rule to a pair $(x,u)$ will yield a walk that
spends equal amounts of
time at each integer $x\!\in\!\{0,\dots,99\}$ on average:
\begin{enumerate}[1.]
    \item $d_1 \sim \D, ~~ d_2 \sim \D, ~~ u \leftarrow \modoneb{u + d_1 - d_2}$
    \item \textbf{if} $u < 0.5$\textbf{:}
    \item \qquad $x \leftarrow \mathrm{mod}(x+1, 100)$
    \item \textbf{else:}
    \item \qquad $x \leftarrow \mathrm{mod}(x-1, 100)$
\end{enumerate}
If the random stream $\D$ gives independent uniform variates, then
the walk will take around $50^2\te2500$
steps to move half way around the ring. However, if adjacent numbers
from the stream $\D$ are very similar, then motion of the state will
persist in a single direction for a long time. For strong
dependencies, the typical time to move half way around the ring is
more like $50$~steps, which is 50~times faster.

There is a similarity here with Hamiltonian Monte Carlo methods
\citeg{neal2011,girolami2011}. In these methods, an auxiliary momentum
variable is included in the state space of the chain. A deterministic
simulation jointly moves the original and auxiliary variables in a way
that reduces random walk behavior of the chain. Hamiltonian dynamics
cannot be simulated exactly. Instead, a discrete approximation is
performed, which does not itself leave the target distribution
invariant and requires correction with a Metropolis step. Standard
Hamiltonian dynamics are only defined for differentiable
log-densities. It is possible that the methods presented
in this paper can use dependencies in random streams to improve mixing
in more general problems. Identifying such practical ways to exploit
dependencies is an interesting future direction of research.

\section{Discussion}

We have shown that \emph{arbitrary} streams of numbers can be used to
drive a Markov chain Monte Carlo algorithm that leaves a target
distribution invariant. If the stream has sufficient variability, or
if a source of ideal random numbers is used occasionally, the
algorithm will also be ergodic. This claim may be counter-intuitive:
what if an adversary set the numbers in the stream?
Given our initial condition, an adversary could make our Markov chain
walk to any state of their choosing, including one with incredibly low
probability. However, if we replaced the initial condition with
another drawn from the target distribution, it is unlikely that
the same stream would
direct us to an atypical state. The stream of numbers must be picked
without looking at the state of our chain, but otherwise it really can
be picked arbitrarily.

This work was originally motivated by theoretical neuroscience
research into Markov chain Monte Carlo implementations in networks of
neurons. The idea that MCMC may be implemented in the brain is present
from early neural network models \citep{alspector1989}, to more recent
models of cognition \citep{ullman2010} and spiking neurons \citep{buesing2011}.
Our theoretical contribution
shows that a source of independent random variates is not required to
construct Markov chains for complex target distributions. Indeed a
system driven by dependent randomness may even perform better. The
particular algorithms suggested here may not be neurally plausible,
but it seems unlikely that \emph{if} MCMC is implemented in the brain,
truly independent variates are used.

\section{Future work}

In certain circumstances,
sequences of pseudo-random numbers can be used to obtain consistent
estimates from Markov chains \citep{chentsov1967}. Recent work has
used \emph{quasi}-random numbers, which are deliberately `more
uniform' than pseudo-random numbers, to drive a Metropolis method
\citep{owen2005}, obtaining lower variance estimation in certain
circumstances. Our work has a different purpose: we wish to use
arbitrary sequences, potentially with any MCMC algorithm, and get an
operator that leaves the target distribution invariant.
In doing so, we can safely interleave our
operators with any other MCMC operators. However, relating our work to
that on quasi-Monte Carlo, and trying to exploit dependencies in
practice (as suggested in section~\ref{sec:advantage}) is an
interesting future direction.

In this work, we have adopted a model for non-uniform random variate
generation that uses continuous random inputs. This model is commonly
adopted, although it is also possible to use random bits as input
instead \citep{devroye1986}. Another interesting future direction
might be attempting to sample from a discrete target distribution
using a stream of discrete dependent variates.

There are several popular software packages for automatically
constructing samplers from user-specified models. Examples include the
long-standing BUGS project \citep{lunn2009}, and more recently, Church
\citep{goodman2008}. Developing our strategy into an automatic code
transformation procedure would allow widespread use of Monte Carlo
methods that require fewer high-quality random numbers. Such
developments may be useful as more large-scale parallel simulations
are deployed.

\section{Conclusion}

For most statistical purposes, current MCMC practice with
pseudo-random number generators works well. This situation can change
with computational paradigms: for example, care is required when
generating many parallel streams of pseudo-random numbers or when
pseudo-random numbers are combined with physical sources of entropy.
This paper
provides a means to make MCMC robust to poor-quality random numbers,
and so provides a new standard to check against. Our practical example
demonstrated a very long run of a deterministic Markov chain, with no
source of randomness at all, that gave useful results. In general it
would be wise to randomize at least some of the steps. However, with
our modifications, the commonly-perceived need for vast numbers of
random variates for MCMC simulations may be illusory.

\section*{Acknowledgments}

We received useful comments from Peter Dayan, Chris Eliasmith, Geoffrey Hinton,
Andriy Mnih, Radford Neal, Vinayak Rao, and Ilya Sutskever.

\let\origurl\url
\renewcommand{\url}[1]{\penalty10000 \hskip.5em
    plus\linewidth \interlinepenalty10000\penalty200
    \hskip-.17em plus-\linewidth minus.11em \origurl{#1}}

\bibliographystyle{Chicago}
\bibliography{bibs}

\end{document}